\documentclass[useAMS,usenatbib,usegraphicx]{mn2e}
\def\msol{\hbox{\kern 0.20em $M_\odot$}}

\newcommand{\lsol}{\hbox{\kern 0.20em $L_\odot$}}
\newcommand{\g}{\hbox{\kern 0.20em g}}
\newcommand{\gmu}{\hbox{\kern 0.20em g$^{-1}$}}
\newcommand{\kg}{\hbox{\kern 0.20em kg}}
\newcommand{\pc}{\hbox{\kern 0.20em pc}}
\newcommand{\mum}{\hbox{\kern 0.20em $\mu$m}}
\newcommand{\mumd}{\hbox{\kern 0.20em $\mu$m$^{-2}$}}
\newcommand{\cm}{\hbox{\kern 0.20em cm}}
\newcommand{\m}{\hbox{\kern 0.20em m}}
\newcommand{\km}{\hbox{\kern 0.20em km}}
\newcommand{\nm}{\hbox{\kern 0.20em nm}}
\newcommand{\s}{\hbox{\kern 0.20em s}}
\newcommand{\h}{\hbox{\kern 0.20em h}}
\newcommand{\smu}{\hbox{\kern 0.20em s$^{-1}$}}
\newcommand{\srmu}{\hbox{\kern 0.20em sr$^{-1}$}}
\newcommand{\smd}{\hbox{\kern 0.20em s$^{-2}$}}
\newcommand{\an}{\hbox{\kern 0.20em an}}
\newcommand{\anmu}{\hbox{\kern 0.20em an$^{-1}$}}
\newcommand{\yr}{\hbox{\kern 0.20em yr}}
\newcommand{\yrmu}{\hbox{\kern 0.20em yr$^{-1}$}}
\newcommand{\Myr}{\hbox{\kern 0.20em Myr}}
\newcommand{\Mymu}{\hbox{\kern 0.20em Myr$^{-1}$}}
\newcommand{\K}{\hbox{\kern 0.20em K}}
\newcommand{\pcmu}{\hbox{\kern 0.20em pc$^{-1}$}}
\newcommand{\pcmd}{\hbox{\kern 0.20em pc$^{-2}$}}
\newcommand{\pcmt}{\hbox{\kern 0.20em pc$^{-3}$}}
\newcommand{\kms}{\hbox{\kern 0.20em km\kern 0.20em s$^{-1}$}}
\newcommand{\kmpd}{\hbox{\kern 0.20em km$^{2}$}}
\newcommand{\kpc}{\hbox{\kern 0.20em kpc}}
\newcommand{\cms}{\hbox{\kern 0.20em cm\kern 0.20em s$^{-1}$}}
\newcommand{\erg}{\hbox{\kern 0.20em erg}}
\newcommand{\ergs}{\hbox{\kern 0.20em erg}}
\newcommand{\cmpd}{\hbox{\kern 0.20em cm$^2$}}
\newcommand{\cmmd}{\hbox{\kern 0.20em cm$^{-2}$}}
\newcommand{\cmms}{\hbox{\kern 0.20em cm$^{-6}$}}
\newcommand{\cmpt}{\hbox{\kern 0.20em cm$^3$}}
\newcommand{\cmmt}{\hbox{\kern 0.20em cm$^{-3}$}}
\newcommand{\mpd}{\hbox{\kern 0.20em m$^2$}}
\newcommand{\mmd}{\hbox{\kern 0.20em m$^{-2}$}}
\newcommand{\mpt}{\hbox{\kern 0.20em m$^3$}}
\newcommand{\mmt}{\hbox{\kern 0.20em m$^{-3}$}}
\newcommand{\mujy}{\hbox{\kern 0.20em $\mu$Jy}}
\newcommand{\mjy}{\hbox{\kern 0.20em mJy}}
\newcommand{\Mj}{\hbox{\kern 0.20em MJy}}
\newcommand{\jy}{\hbox{\kern 0.20em Jy}}
\newcommand{\ghz}{\hbox{\kern 0.20em GHz}}
\newcommand{\G}{\hbox{\kern 0.20em G}}
\newcommand{\muG}{\hbox{\kern 0.20em $\mu$G}}

\title[The density structure of the L1157 molecular outflow]{The density structure of the L1157 molecular outflow}

\author[G\'omez-Ruiz et al.]{A.I. G\'omez-Ruiz$^{1,} $\thanks{Current address: Instituto Nacional de Astrof\'isica, \'Optica y Electr\'onica (INAOE), Luis Enrique Erro No.1, C.P. 72840, Tonantzintla, Puebla, Mexico. e-mail: aigomez@inaoep.mx}, C. Codella$^{1}$, B. Lefloch$^{2,3}$, M. Benedettini$^{4}$, G. Busquet$^{5,4}$, \newauthor C. Ceccarelli$^{2,3}$, B. Nisini$^{6}$, L. Podio$^{1}$, S. Viti$^{7}$
\\
\\
$^{1}$ INAF, Osservatorio Astrofisico di Arcetri, Largo E. Fermi 5, 50125 Firenze, Italy \\
$^{2}$ Univ. Grenoble Alpes, IPAG, F-38000 Grenoble, France \\
$^{3}$ CNRS, IPAG, F-38000 Grenoble, France\\
$^{4}$ INAF, Istituto di Astrofisica e Planetologia Spaziali, via Fosso del Cavaliere 100, 00133, Roma, Italy \\
$^{5}$ Instituto de Astrof\'isica de Andaluc\'ia, CSIC, Glorieta de la Astronom\'ia s/n, E-18008 Granada, Spain \\
$^{6}$ INAF, Osservatorio Astronomico di Roma, via di Frascati 33, 00040, Monte Porzio Catone, Italy \\
$^{7}$  Department of Physics and Astronomy, University College London, London, UK \\
}

\begin{document}

\date{Accepted date. Received date; in original form date}

\pagerange{\pageref{firstpage}--\pageref{lastpage}} \pubyear{2011}

\maketitle

\label{firstpage}

\begin{abstract}
We present a multiline CS survey towards the brightest bow-shock B1 in the prototypical chemically active protostellar outflow L1157. We made use of (sub-)mm data obtained in the framework of the Chemical HErschel Surveys of Star forming regions (CHESS) and Astrochemical Surveys at IRAM (ASAI) key science programs. We detected $^{12}$C$^{32}$S, $^{12}$C$^{34}$S, $^{13}$C$^{32}$S, and $^{12}$C$^{33}$S emissions, for a total of 18 transitions, with $E_{\rm u}$ up to $\sim$ 180 K. The unprecedented sensitivity of the survey allows us to carefully analyse the line profiles, revealing high-velocity emission, up to 20 km s$^{-1}$ with respect to the systemic. The profiles can be well fitted by a combination of two exponential laws that are remarkably similar to what previously found using CO. These components have been related to the cavity walls produced by the $\sim$ 2000 yr B1 shock and the older ($\sim$ 4000 yr) B2 shock, respectively. The combination of low- and high-excitation CS emission was used to properly sample the different physical components expected in a shocked region. Our CS observations show that this molecule is highlighting the dense, $n_{\rm H_2}$ = 1--5 $\times$ 10$^{5}$ cm$^{-3}$, cavity walls produced by the episodic outflow in L1157. In addition, the  highest excitation (E$_u$ $\geq$ 130 K) CS lines provide us with the signature of denser (1--5 $\times$ 10$^{6}$ cm$^{-3}$) gas, associated with a molecular reformation zone of a dissociative J-type shock, which is  expected to arise where the precessing jet impacting the molecular cavities. The CS fractional abundance increases up to $\sim$ 10$^{-7}$ in all the kinematical components. This value is consistent with what previously found for prototypical protostars and it is in agreement with the prediction of the abundances obtained via the chemical code Astrochem.
\end{abstract}


   \maketitle

    \begin{figure}
   \centering
      \includegraphics[angle=0,width=8.5cm]{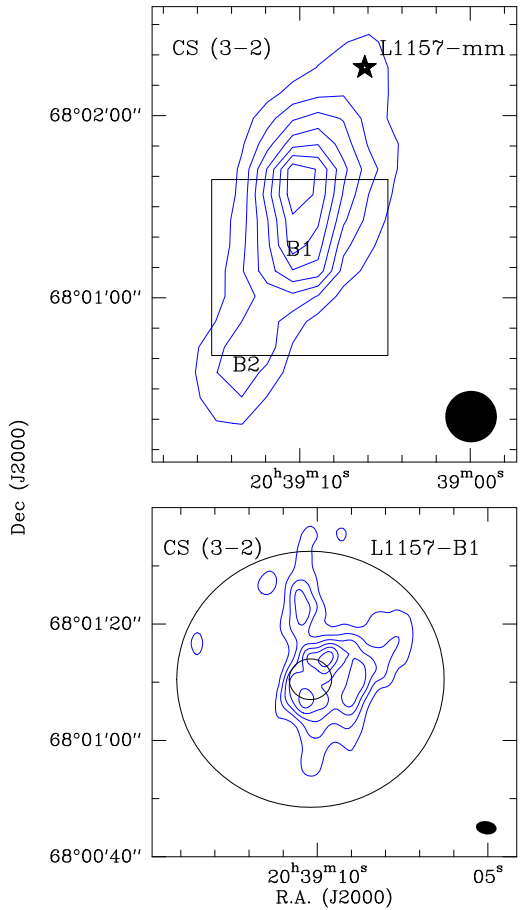}
      \caption{The CS(3--2) emission integrated from $\sim$ $-$21 to $+$5 km s$^{-1}$ in the blue-lobe of the L1157 outflow. Upper panel: IRAM 30-m single-dish map of the blue lobe (first contour and steps of 3 K km s$^{-1}$, corresponding to $\sim$ 3$\sigma$). The star points out the position of the driving protostar L1157-mm. The labels indicate the main blue-shifted knots B1 and B2. The ellipse in the bottom-right corner shows the HPBW. Lower panel: IRAM PdBI CS(3--2) image (the synthesized beam is reported in the bottom-right corner) of the B1 clump (Benedettini et al. 2013; first contour and steps of 0.5 Jy beam$^{-1}$ km s$^{-1}$, corresponding to $\sim$ 3$\sigma$). Circles are for the smallest (7$\arcsec$) and largest (44$\arcsec$) HPBW of the present dataset.
              }
         \label{cs-30m-beam}
   \end{figure}

\section{Introduction}

Bipolar fast jets driven by protostars are present during the earliest stages of low-mass star formation. The interaction of the jets with the high-density ambient medium creates shocks, which in turn trigger endothermic chemical reactions and ice grain mantle sublimation or sputtering. The chemical composition of the gas phase is consequently altered and enriched. One of the best studied protostellar outflows with strong shocks and very active chemistry is the bipolar outflow driven by the low-mass Class 0 protostar L1157-mm, located at a distance of $\simeq$ 250 pc  \citep{Loon07}, and with a luminosity of $L_{\rm bol} \sim$ 3 $L_{\odot}$ { \citep{Tobin10}. Several blue- and red-shifted shocks have been revealed using CO (Gueth et al. 1996, 1998) and H$_2$ \citep{Neufeld09,Nisini10}. The interferometric CO maps made of this outflow showed that the spatial-kinematic structure of the blue lobe is reproduced by a model of two limb-brightened cavities with slightly different axes \citep{Gueth96}. In particular, the shocks that produced the two cavities in the blue-lobe were labelled B2 and B1 (with kinematical ages of $\sim$ 4000 and 2000 yr, respectively), of which B1 has been found to dominate the emission of several molecular transitions \citep{Gueth98,Zhang95,Zhang2000,Tafalla95,Bachiller2001}. Interferometric observations have revealed the complex structure of the B1 shock, consisting of multiple clumps \citep{Bene07,Codella09,GR13} and characterized by an east-west chemical stratification \citep{Bene07,Bene13}. The B1 and B2 positions of the L1157 outflow have been the target of multiple molecular line studies that establish L1157 as the prototype of chemically active outflows \citep{Bachiller97,Bachiller2001}.

Under the framework of the IRAM Large Program ASAI\footnote{http://www.oan.es/asai} (Astrochemical Surveys at IRAM) and the Herschel key project CHESS \footnote{http://www-laog.obs.ujf-grenoble.fr/heberges/chess/} (Chemical HErschel Surveys of Star forming regions), an unbiased spectral line survey of the L1157-B1 position has been carried out in the range 80-350 GHz (with the IRAM-30m antenna) and 500--2000 GHz (with \emph{Herschel}-HIFI).
The first results of the CHESS program \citep{Lefloch10,Lefloch12,Codella10,Codella12a,Codella12b,Codella13,Bene12,Busquet14} confirmed the chemical richness of the shocked gas, showing bright emission due to species released by grain mantles, such as NH$_3$, H$_2$O, and CH$_3$OH. In addition, we found that all the CO lines profiles are well fit by a linear combination of three exponential laws of the form $I(v)= exp{-|v/v_0|}$, which trace three different kinematical and thermal components. These components are (i) a molecular reformation region of a dissociative J-shock occurring at the B1 position (also refereed to $g_1$; Lefloch et al. 2012), (ii) the B1 cavity wall ($g_2$), and (iii) the older B2 cavity ($g_3$).
For each component the CO molecule has provided severe constrains of the kinetic temperature, T$_k \sim$ 210, 64, and 23 K, respectively, whereas only lower limits of the gas density have been provided.

\begin{table}
\centering
\caption{CS critical densities, $n_{\rm cr}$, for typical kinetic temperature of 60 K.}
\label{tab:ncrit}
\centering
\begin{tabular}{lccc}
\hline
\hline
Transition & A$_{\rm ij}$ (\smu) & C$_{\rm ij}$ (\cmpt \smu) & $n_{\rm cr}$ (\cmmt) \\
\hline
2--1    & 1.679E-05 & 4.35E-11 & 3.85E+05  \\
3--2    & 6.071E-05 & 5.00E-11 & 1.21E+06  \\
5--4    & 2.981E-04 & 5.42E-11 & 5.50E+06  \\
6--5    & 5.230E-04 & 5.52E-11 & 9.47E+06  \\
7--6    & 8.395E-04 & 5.64E-11 & 1.48E+07  \\
10--9   & 2.496E-03 & 6.15E-11 & 4.05E+07  \\
11--10  & 3.336E-03 & 6.35E-11 & 5.25E+07  \\
12--11  & 4.346E-03 & 6.58E-11 & 6.60E+07  \\
\hline
\hline
\end{tabular}
\begin{center}
Note-- A$_{\rm ij}$ coefficients and collisional rates (C$_{\rm ij}$) from LAMDA data base (http://home.strw.leidenuniv.nl/$\sim$moldata/CS.html). The C$_{\rm ij}$ are for CS-H2, scaled from CS-He (Lique et al. 2006).  \\
\end{center}
\end{table}

The CS molecule can fix the density uncertainty, given (i) it is a standard density tracer \citep{Tak07},
 due to the high critical density of most of the transitions (see Table \ref{tab:ncrit}), i.e. subthermal excitation, and (ii) the CS abundance, $X$(CS), in shocked material along molecular
outflows \citep[e.g.,][]{Bachiller2001,Wakelam05,Tafalla10} increases up to an
order of magnitude with respect to what observed in quiescent clouds.
Indeed CS can be efficiently formed
from species such as OCS, that together with H$_2$S is the most abundant S-bearing molecule
released from the grain mantles after a shock, and injected into the gas
phase (Wakelam et al. 2004, 2005, Codella et al. 2005).
This hypothesis is supported by recent results on molecular ions by \citet{Podio14}, who found that
if OCS is released from dust grains it is possible to simultaneously reproduce the CS and HCS$^+$ abundances.
The CS emission can thus be used to characterize the physical conditions of the gas components
in L1157-B1 as revealed by CO.
Only the combination of low$-J$ CS lines observed with IRAM and the high$-J$ CS lines observed with \emph{Herschel}-HIFI
can properly sample the different physical components that may compose the line profiles. The observations are reported in Sect. 2,
while in Sect. 3 we describe the line profiles and their decomposition into different components; the physical
conditions and the derived CS abundances are reported in Sect. 4. Our conclusions are summarized in Sect. 5.

\begin{table*}
\centering
\caption{Transitions, parameters, and integrated intensities of the CS and isotopologues lines observed.}
\label{nspecie}
\centering
\begin{tabular}{lccccccccc}
\hline
\hline
Transition$^{\rm (a)}$ & $\nu_{\rm 0}$ & $E_{\rm u}$ & Telescope & HPBW & $B_{\rm eff}$ & $F_{\rm eff}$ & $T_{\rm MB}^{\rm peak} $ $^{\rm (b)}$ & $V_{\rm min},V_{\rm max}$ & $\int$ $T_{\rm MB}dv^{\rm (d)}$ \\
           & (GHz)            & (K) & & ($''$) & &  & (K) & (km s$^{-1}$) &(K km s$^{-1}$)\\
\hline
\multicolumn{10}{c}{$^{12}$C$^{32}$S}\\
\hline
2--1     & 97.98095       & 7     & IRAM & 26 & 0.80 & 0.95 & 2.45(0.002) & --19, 6 & 17.28\\
3--2     & 146.96903      & 14    & IRAM & 17 & 0.74 & 0.93 & 2.96(0.004) & --19, 6 & 21.03\\
5--4     & 244.93556      & 35    & IRAM & 10 & 0.56 & 0.94 & 2.34(0.004) & --16, 6 & 17.62\\
6--5     & 293.91209      & 49    & IRAM & 9  & 0.45 & 0.88 & 1.61(0.006) & --16, 6 & 12.67\\
7--6     & 342.88285      & 65    & IRAM & 7  & 0.35 & 0.82 & 0.96(0.026) & --16, 3 &  7.40\\
10--9    & 489.75104      & 129   & HIFI & 44 & 0.73 & 0.96 & 0.06(0.008) & --4, 3  &  0.38\\
11--10   & 538.68900      & 155   & HIFI & 39 & 0.73 & 0.96 & 0.04(0.006) & --5, 3  &  0.34\\
12--11   & 587.61649      & 183   & HIFI & 36 & 0.72 & 0.96 & 0.03(0.009) & --3, 4  &  0.25\\
\hline
\multicolumn{10}{c}{$^{12}$C$^{34}$S}\\
\hline
2--1     &  96.41295      & 7  & IRAM & 26  & 0.80 & 0.95 & 0.15(0.001) & --14, 4 & 1.12\\
3--2     & 144.61710      & 14 & IRAM & 17  & 0.74 & 0.93 & 0.16(0.002) & --13, 4 & 1.16\\
5--4     & 241.01608      & 35 & IRAM & 10  & 0.57 & 0.94 & 0.11(0.003) & --10, 4 & 0.77\\
6--5     & 289.20907      & 49 & IRAM & 9   & 0.46 & 0.88 & 0.05(0.007) & --7, 3  & 0.39\\
7--6     & 337.39645      & 50 & IRAM & 7  & 0.35 & 0.82 & $<$0.08$^{\rm (c)}$ & - &  -\\
10--9    & 481.91586      & 96  & HIFI & 44 & 0.73 & 0.96 & $<$0.05$^{\rm (c)}$ & -  &  -\\
11--10   & 530.07122      & 115 & HIFI & 40 & 0.73 & 0.96 & $<$0.05$^{\rm (c)}$ & -  &  -\\
12--11   & 578.21605      & 135 &HIFI & 37 & 0.72 & 0.96 & $<$0.05$^{\rm (c)}$ & -  &  -\\
\hline
\multicolumn{10}{c}{$^{13}$C$^{32}$S}\\
\hline
2--1     & 92.49430       & 7   & IRAM & 26  & 0.80 & 0.95 & 0.05(0.002) & --15, 5 & 0.34\\
3--2     & 138.73933      & 13  & IRAM & 17  & 0.74 & 0.93 & 0.06(0.003) & --11, 4 & 0.43\\
5--4     & 231.22099      & 33  & IRAM & 10  & 0.58 & 0.94 & 0.03(0.003) & --10, 4 & 0.25\\
6--5     & 277.45540      & 47  & IRAM & 9   & 0.46 & 0.88 & $<$0.02$^{\rm (c)}$ & -  & -\\
11--10   & 508.52814      & 146 & HIFI & 42 & 0.73 & 0.96 & $<$0.03$^{\rm (c)}$ & -  &  -\\
\hline
\multicolumn{10}{c}{$^{12}$C$^{33}$S}\\
\hline
2--1     & 97.17206       & 6    & IRAM & 26  & 0.80 & 0.95 & 0.02(0.002) & --11, 4 & 0.15\\
3--2     & 145.15553      & 12   & IRAM & 17  & 0.74 & 0.93 & 0.03(0.003) & --7, 4  & 0.25\\
5--4     & 242.91361      & 28   & IRAM & 10  & 0.56 & 0.94 & 0.02(0.004) & --7, 3  & 0.12\\
6--5     & 291.48593      & 39   & IRAM & 9   & 0.46 & 0.88 & $<$0.02$^{\rm (c)}$ & -  & -\\
\hline
\hline
\end{tabular}
\begin{center}
$^{\rm (a)}$ Transition properties are taken from the Cologne Database for Molecular Spectroscopy: \citet{muller05}. $^{\rm (b)}$ In parenthesis the r.m.s ($\sigma$) per channel width ($\Delta v$) of 1.4 km s$^{-1}$, except for the 6--5 lines with $\Delta v\sim$ 2.1 km s$^{-1}$. $^{\rm (c)}$ 3$\sigma$ upper limit. $^{\rm (d)}$ The integrated area between $V_{\rm min},V_{\rm max}$ (see Sect. 3.1).  \\
\end{center}
\end{table*}

   \begin{figure}
   \centering
      \includegraphics[bb=85 30 340 740,angle=0,width=6.4cm]{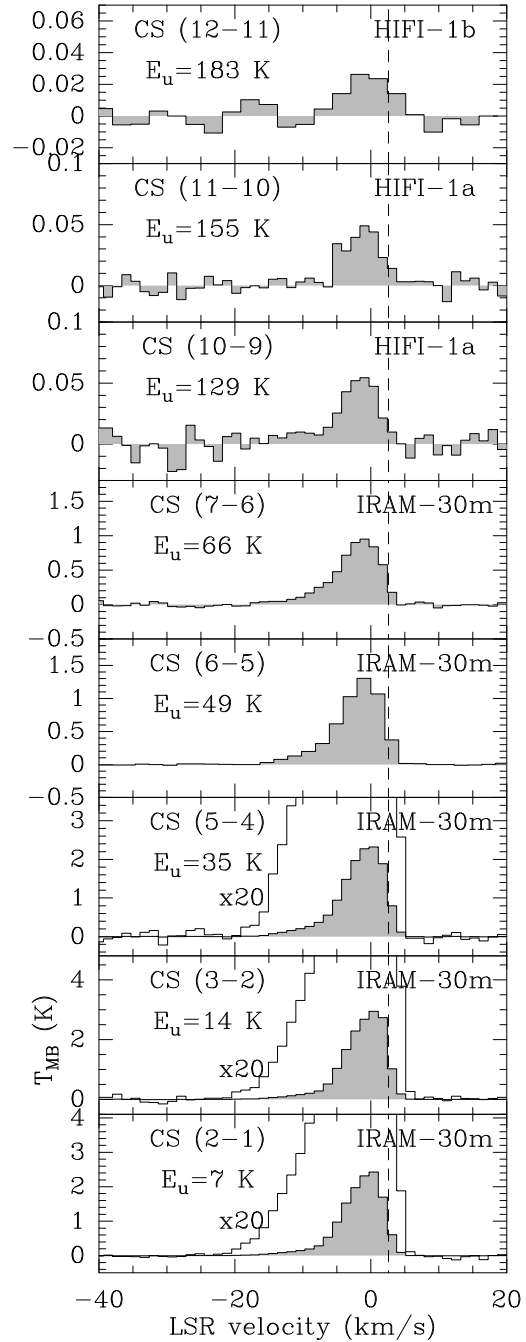}
      \caption{Spectra of the CS main isotope transitions detected by IRAM-30m and \emph{Herschel}-HIFI at L1157-B1. Indicated by labels are the corresponding rotational transition, the upper level energy of the transition ($E_{\rm u}$), and the instrument.The three lowest panels also show a zoom in of the wings. The dashed vertical line indicates the cloud V$_{sys}$ of $+$2.6 km s$^{-1}$ \citep{Bachiller97}.
              }
         \label{cs-spec}
   \end{figure}

\section{Observations}

\subsection{\emph{Herschel}-HIFI}
The CS(10--9), (11--10), (12--11) transitions were observed with \emph{Herschel}-HIFI on 2010 October 27 and 2009 August 1 (Obs\_ID 1342207575 and 1342181160), during the unbiased spectral survey CHESS with the HIFI bands 1a and 1b, at the position of the B1 shock in L1157 (see Fig. 1). The pointed coordinates were $\alpha_{\rm 2000}$ = 20$^{\rm h}$ 39$^{\rm m}$ 10$\fs$2, $\delta_{\rm J2000}$ = +68$\degr$ 01$\arcmin$ 10$\farcs$5, i.e. at $\Delta\alpha$ = +25$\farcs$6 and $\Delta\delta$ = --63$\farcs$5 from the driving protostar. The receiver was tuned in double side band mode, with a total integration time of 8912 and 8400 sec to cover bands 1a and 1b, respectively. The Wide Band Spectrometer (WBS) was used, with a velocity resolution of 0.15--0.17 km s$^{-1}$, depending on frequency. All HIFI spectra were smoothed to a common velocity resolution of 1.4 km s$^{-1}$, in order to be compared with the IRAM-30m spectra (see below). The forward ($F_{\rm eff}$) and beam ($B_{\rm eff}$) efficiencies, as well as the HPBW were taken according to Roelfsema et al. (2012) and reported in Table \ref{nspecie}\footnote{The data presented here do not include the updated HIFI calibration values as of September 26, 2014. However, this does not affect significantly the main results of our analysis.}. The \emph{Herschel} data were processed with the ESA-supported package HIPE 8.10\footnote{HIPE is a joint development by the Herschel Science Ground Segment Consortium, consisting of ESA, the NASA Herschel Science Center, and the HIFI, PACS and SPIRE consortia.} \citep[Herschel Interactive Processing Environment:][]{Ott10}. FITS files from level 2 were then created and transformed into GILDAS\footnote{http://iram.fr/IRAMFR/PDB/gildas/gildas.html} format for data analysis. The CS (13--12) at 636532.454 MHz falls right at the outer edge of band 1b  and was missed by our HIFI/CHESS survey. The CS (14--13) at 685435.917 MHz lies in band 2a and was observed  (Obs\_ID:1342207607), but not detected down to an rms of 3.6 mK [Ta*] in a velocity interval of 5 km s$^{-1}$. Therefore, the HIFI/CHESS survey detected the CS transitions from $J_{up}= 10$ up to the $J_{up}=12$.

\subsection{IRAM-30m}
The lower$-J$ CS transitions ($J_{\rm up}=$ 2, 3, 5, 6, 7) were obtained during the IRAM-30m unbiased spectral survey of L1157-B1 (Lefloch et al., in preparation) as part of the ASAI Large Program towards the same position observed with \emph{Herschel}-HIFI. The survey was performed during several runs in 2011 and 2012, using the broadband EMIR receivers, the Fourier Transform Spectrometer (FTS; velocity resolution up to 1.4 km s$^{-1}$) and the WILMA spectrometer (only the 6--5 transitions; $\sim$ 2.1 km s$^{-1}$). All the spectra taken with the FTS were smoothed to a common velocity resolution of 1.4 km s$^{-1}$. The forward and beam efficiencies, as well as the HPBW are reported in Table \ref{nspecie}. As complementary data, the CS(3--2) line emission was mapped at the Nyquist spatial frequency (i.e. every 8$\arcsec$) in a region of $\sim$ 200$\arcsec$ $\times$ 400$\arcsec$, covering  the entire blue lobe of the L1157 outflow in 2011 September with the IRAM-30m antenna (see Fig. 1, upper panel). The spectral resolution was 0.4 km s$^{-1}$. The average rms of the map was $\sim$ 0.1 K per velocity interval.

The total frequency coverage of the survey allow us to observe most of the CS rotational transitions from $J$ = 2--1 up to $J$ = 12--11, as well as some isotopic species transitions (in particular low$-J$ transitions). We did not observe the $J$ = 1--0, 4--3, 8--7, and 9--8 transitions. The spectra in this paper are reported in units of main-beam brightness temperature (T$_{\rm MB}=$T$_A^* \times$ $F_{\rm eff}$/$B_{\rm eff}$), for which we have used the $F_{\rm eff}$ and $B_{\rm eff}$ in Table \ref{nspecie}. Nominal flux calibration uncertainties of 10\% and 20\%, for Herschel and IRAM-30m, respectively, are considered in the analysis.

\section{Results}

\subsection{Detected transitions and line opacities}

The observed transitions of CS and its isotopologues are listed in Table \ref{nspecie}. In total 18 transitions from $^{12}$CS, $^{13}$CS, C$^{33}$S, and C$^{34}$S were detected. In Table \ref{nspecie} are reported the peak intensities (T$_{\rm MB}^{\rm peak}$), the velocity limits of the detection ($V_{min}$ and $ V_{max}$, defined at a 3$\sigma$ detection limit), and the integrated emission within this limits ($\int$ $T_{\rm MB}dv$).
Figures \ref{cs-spec} and \ref{cs-iso-spec} show the spectra of the CS and its isotopologues transitions, respectively.
The strongest lines are CS (2--1) and (3--2), with typical peak intensities between $\sim$ 2-3 K.
The highest velocities detected in these transitions reach up to V$_{LSR}$$\sim$ $-$20 km s$^{-1}$
\citep[cloud velocity is V$_{sys}$ = $+$2.6 km s$^{-1}$: e.g.][]{Bachiller97}.
On the other hand, the highest frequency lines observed with HIFI have typical peak intensities of $\sim$ 30-60 mK,
and they span a much narrow velocity range, with the highest velocity up to V$_{LSR}$$\sim$ $-$5 km s$^{-1}$.
The isotopologues were detected up to the $J$ = 6--5 transition, with peak intensities in the 20-160 mK range.

  \begin{figure}
   \centering
      \includegraphics[bb=51 220 507 738,angle=0,width=9.0cm]{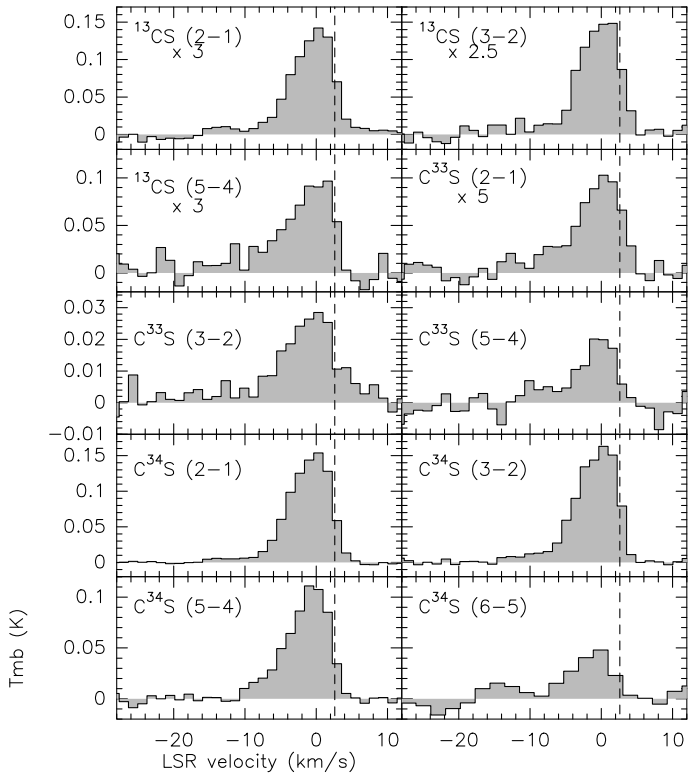}
      \caption{Spectra of the CS isotopologues transitions detected by IRAM-30m at L1157-B1 (in T$_{MB}$ scale). Each panel indicates the corresponding rotational transition. The vertical dashed line indicates the cloud velocity (V$_{sys}=+$2.6 km s$^{-1}$).
              }
         \label{cs-iso-spec}
   \end{figure}

Since only the 2--1 line of all the isotopologues was detected in a wide range of velocities with a
good signal-to-noise ratio (i.e. $>$ 4), we use it to determine the CS (2--1) line opacity as a function of velocity. For this aim we assumed the following abundance ratios: $^{12}$C/$^{13}$C=75, $^{32}$S$^{34}$S=22, $^{32}$S/$^{33}$S=138 \citep{Wilson94,Chin96}. Figure \ref{cs-opac} shows the opacity the CS (2--1) transition as a function of velocity, obtained from the C$^{32}$S/C$^{34}$S and $^{12}$CS/$^{13}$CS ratios. We found $\tau$$\sim$0.05 at V$_{LSR}$ of $-$7.5 km s$^{-1}$ (the highest velocity in which the line ratio is $>$ 2$\sigma$), while $\tau$$\sim$1 at the cloud velocity. With these information we conclude that the CS emission is at most moderately optically thick (i.e. $\tau <$ 2), and at the outflow velocities (V$_{LSR} < -$6 km s$^{-1}$) there is evidence for optically thin CS emission. Figure \ref{cs-opac} also shows the excitation temperature ($T_{\rm ex}$--$T_{\rm bg}$), uncorrected for beam filling (ff), and as a function of velocity (discussed in Sect. 4).

   \begin{figure}
   \centering
      \includegraphics[angle=-90,width=9.2cm]{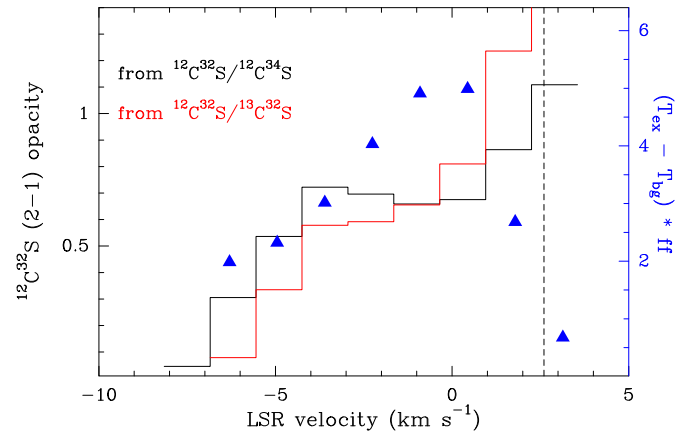}
      \caption{The CS opacity as calculated from the C$^{32}$S/C$^{34}$S (2--1) and $^{12}$CS/$^{13}$CS (2--1) line ratios (black and red histograms, respectively). Blue triangles show the excitation temperature calculated from opacity (not corrected for beam filling factor, i.e. (T$_{ex}$--T$_{bg}$) $\times$ ff). The dashed line indicates the cloud velocity as in Fig. 2 (V$_{sys}=+$2.6 km s$^{-1}$).
              }
         \label{cs-opac}
   \end{figure}

The IRAM-30m CS (3--2) map integrated between V$_{LSR}$ of $-$21 to $+$5 km s$^{-1}$, shown in Fig. 1, traces the large scale CS gas structure of B1 and B2, being consistent with previous results by \citet{Bachiller2001}. As shown by \citet{Bene13} the interferometric data recovers most of the emission at the highest negative velocities (100\% at V$_{LSR} < - 6$ km s$^{-1}$), while it looses a considerable fraction of the emission ($\sim$ 40\%) around the cloud velocity (see Benedettini et al. 2013 for details). Thus, the total integrated emission observed in the interferometric map shown in the lower panel of Fig. 1 is dominated by the high-velocity structures. As shown below, the information provided by this map will allow us to constrain the physical conditions obtained from our multiline CS analysis.

\subsection{CS spectral line components}

Thanks to the high signal-to-noise ratio of the ASAI data, we have analysed the profiles of the CS transitions from $J$=2--1 up to  $J$=7--6 following the approach of Lefloch et al. (2012). Figure \ref{cs-exp} shows the line profiles of the CS transitions $J$=2--1 up to $J$=7--6 on a linear-logarithmic scale. We identify two physical components, whose intensity-velocity distributions can be fit with an exponential law $\rm I(v) \propto exp(-|v/v_0|)$ with the same slope at all $J$, but differing relative intensities.

For $J$=6--5 and $J$=7--6, the intensity-velocity distribution in the line profiles is well fit by a single exponential law of the form $\rm I(v) \propto exp(-|v/v_0|)$, with $v_0= 4.4\kms$ for velocities in the range $\approx -18$ to $0\kms$. This component dominates the emission between $-$20 and $-8\kms$ in the other CS transitions.  Its contribution to  each  transition can be obtained from a simple scaling to the $J$=7--6 line profiles. After removing the contribution of this component, an emission excess is observed in the $J$=2--1, $J$=3--2 and $J$=5--4 transitions, which is detected only at velocities between $-8\kms$ and $+2\kms$. As shown  in Fig. \ref{cs-exp-g3}, the emission excess is well fitted by a second exponential law $\rm I(v) \propto exp(-|v/2.5|)$, hence with an exponent different from that of the first CS component. This second component contributes to about half of the total flux emitted in the $J$=2--1 and $J$=3--2 (see sect. 4 and Table 3). The lower panels in Fig. \ref{cs-exp} show the simultaneous two components fit to the line 2--1, 3--2, and 5--4 profiles. Important to note is that the crossing point between the two components shift back and forth from the 2--1 to the 5--4 transition. The latter can in part be due to the higher uncertainties in the fitting of the emission excess, in which we usually have fewer channels than for the first component (see Fig. \ref{cs-exp-g3}) to make the fit. The uncertainty of the fitted ordinate of this second component is up to $\sim$ 23\% and therefore close to the intensity contrast between the 2--1 and 3--2 emission of this component (3--2 is about 28\% stronger than 2--1). Alternatively, this behaviour can also be due to excitation, in this case with the emission peak of the second component in the 3--2 transition. Finally, weak contamination from the reference position may understimate the emission close to the ambient velocity, contributing to the apparent shift. Unfortunatelly, with the present data it is not possible to disentangle between these possibilities. However, due to our method to obtain the fluxes of this second component (sect. 4), the above mentioned uncertainty does not affect significantly the analysis of its emission.


The high signal-to-noise ratio of the line profiles of the C$^{34}$S $J$=2--1, $J$=3--2 and $J$=5--4 permits a similar analysis of the intensity-velocity distribution. Like for the high$-J$ CS transitions, we find that the  C$^{34}$S (5--4) line profile is well fitted by one single exponential law $\rm I(v) \propto exp(-|v/4.4|)$. An emission excess is observed between -8 and $+2\kms$ in the lower-J transitions, which is well fitted by the second exponential law $\rm I(v) \propto exp(-|v/2.5|)$. Like for CS, the spectral slope of both components in C$^{34}$S is independent of the transitions considered, which implies excitation conditions independent of the velocity. Therefore, the profile analysis of C$^{34}$S yields results similar to CS.

   \begin{figure}
   \centering
\includegraphics[bb=50 30 350 755,angle=0,width=7.0cm]{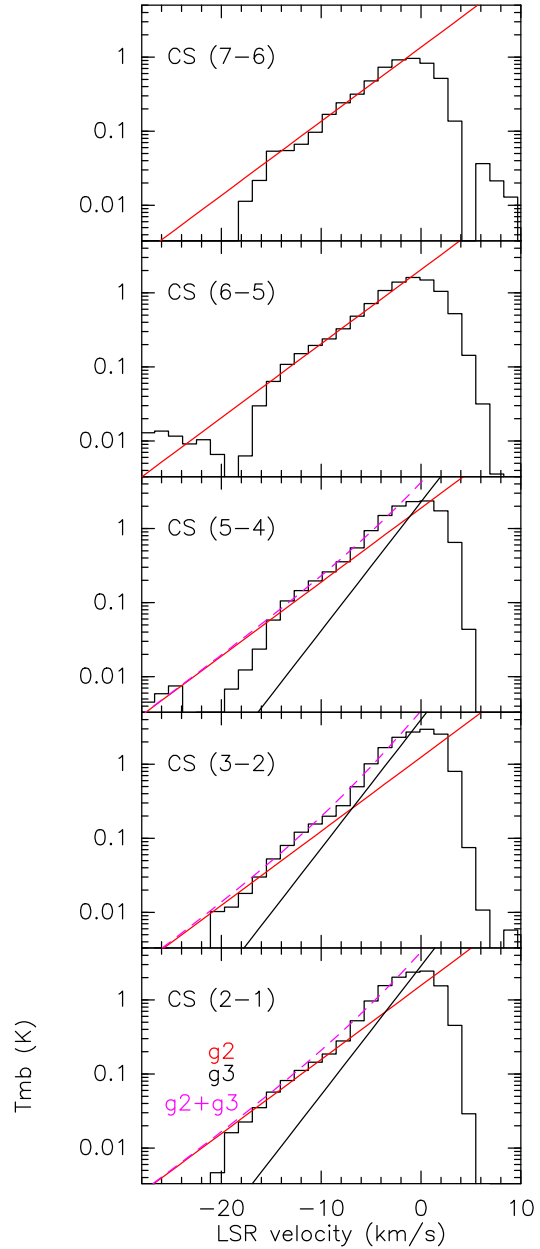}
      \caption{Exponential fits to the CS transitions. The components $\rm I(v) \propto exp(-|v/4.4|)$ and $\rm I(v) \propto exp(-|v/2.5|)$ are shown by the red and black lines, respectively; while their linear combination is indicated by the dashed magenta line. These exponential components correspond to the $g_2$ and $g_3$ components, respectively, found by Lefloch et al. (2012) in the CO emission.
              }
         \label{cs-exp}
   \end{figure}

   \begin{figure}
   \centering
\includegraphics[angle=0,width=8.0cm]{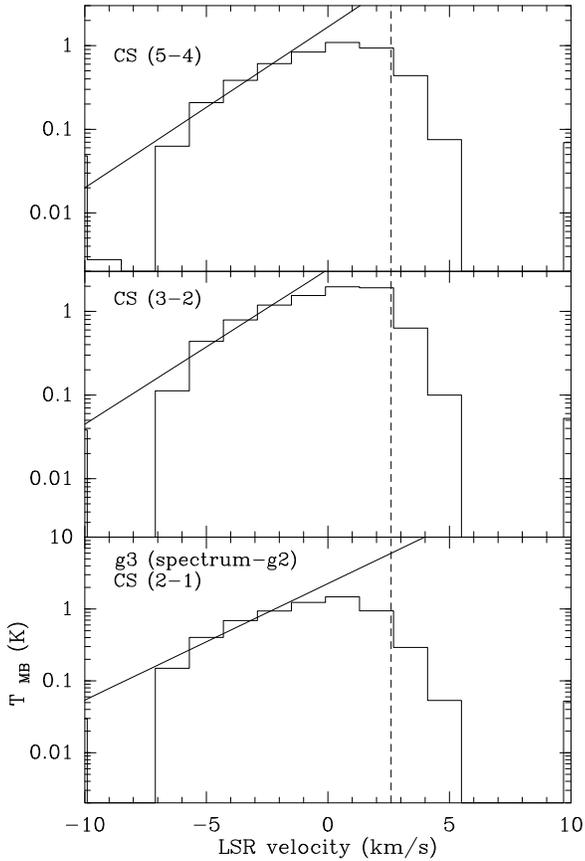}
      \caption{The emission excess, after removing the emission from the first component ($\rm \propto exp(-|v/4.4|)$). The black line shows the fit of a second exponential component $\rm I(v) \propto exp(-|v/2.5|)$.
              }
         \label{cs-exp-g3}
   \end{figure}

Because of the lower signal-to-noise of the data, it was not possible to analyse the intensity-velocity distribution of the CS transitions observed in the HIFI range. As discussed in Sect. 4, the emission of these transitions arises from a region of higher excitation than the $\rm I(v) \propto exp(-|v/4.4|)$ and $\rm I(v) \propto exp(-|v/2.5|)$ components observed with the IRAM 30m telescope.



\subsection{Origin of the emission}

In Fig. \ref{log-pdb-30m} we compare the CS (3--2) profile obtained at the IRAM 30-m telescope with that obtained at the PdBI (Benedettini et al. 2013) when integrating the emission over the region covered by the single-dish main-beam. While the IRAM-30m profile is fitted, as discussed above, with a linear combination of the two exponential components, the spectrum derived from the interferometer is well fitted by the first $\rm \propto exp(-|v/4.4|)$ component only (convolving the same PdBI map to an even larger beam, Benedettini et al. 2013 also found the first component dominating the 3--2 spectrum). Since most of the high-velocity emission is recovered by the interferometer (see Sect. 3.1), we conclude that the latter map traces the whole emission from the first CS component, and that this emission arises from the walls of the B1 cavity. The size of the first CS component is  $\simeq  18\arcsec$, as derived from the PdBI image. We conclude that the low-velocity emission, associated with the $\rm I(v) \propto exp(-|v/2.5|)$ component is filtered out by the interferometer, suggesting a larger size for this component.

Furthermore, the spectral signatures of the two CS components are found similar (i.e same slope) to those of the components $g_2$ and $g_3$ detected in CO by  Lefloch et al. (2012), which these authors associated to the B1 and B2 cavities, respectively. As a conclusion, the emission of the dense gas, as traced by CS and its isotopomers, arise from the outflow cavities associated with B1 and B2. For the sake of consistency with our previous work, in what follows, we will refer to the first and second CS component, respectively, as $g_2$ and $g_3$. Also, for consistency with the CO analysis, we will adopt a typical size of $25\arcsec$ for $g_3$.

\begin{figure*}
   \centering
      \includegraphics[angle=-90,bb= 368 17 544 508,width=18cm]{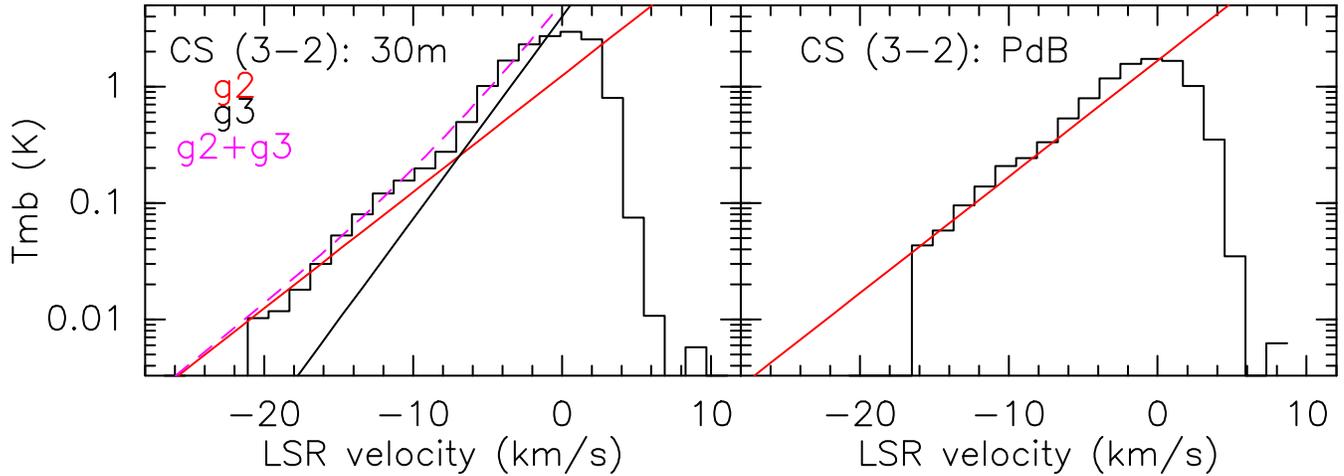}
      \caption{A comparison between the CS (3--2) line profiles at the B1 position, in logarithm scale, as observed with the IRAM-30m (left) and the IRAM-PdBI (right). The PdBI data was convolved to the IRAM-30m angular resolution (17$''$). Both panels also show the exponential components that better fit the line profiles.
Note how the interferometric data is well fitted with the $g_2$ component only, while the single-dish data needs
both $g_2$ and $g_3$ components.}
         \label{log-pdb-30m}
   \end{figure*}

\begin{table}
\centering
\caption{CS velocity-integrated intensities of the components $g_1$, $g_2$, and $g_3$, used in the LVG calculations.}
\label{tab:flux-g2g3}
\centering
\begin{tabular}{lccccccc}
\hline
\hline
Line  & HPBW & \multicolumn{6}{c}{$\int$ $T_{\rm MB}dv$ (K km s$^{-1}$)$^{\rm (a)}$} \\
\cline{3-8}
         & ($''$) & $g_1^{\rm (b)}$ & $g_2^{\rm (b)}$ & $g_3^{\rm (b)}$ & $g_1^{\rm (c)}$ & $g_2^{\rm (c)}$ & $g_3^{\rm (c)}$ \\
\hline
\multicolumn{8}{c}{$^{12}$C$^{32}$S}\\
\hline
2--1     & 26 & - & 8.7  & 8.4 & - & 10.7 & 9.9 \\
3--2     & 17 & - & 9.8  & 11.7 & - & 8.8  & 11.2 \\
5--4     & 10 & - & 13.1 & 4.9 & - & 6.0 & 3.5 \\
6--5     & 9  & - & 9.9 &  -  & -  &  4.3 & - \\
7--6     & 7  & - & 8.1 &  -  & -  &  3.1 & -  \\
10--9    & 44 & 0.39  &  -  & -  & 0.64  & - & - \\
11--10   & 39 & 0.33 &  -  & -  & 0.51  & - & - \\
12--11   & 36 & 0.28  &  -  & -  & 0.43  & - & - \\
\hline
\multicolumn{8}{c}{$^{12}$C$^{34}$S}\\
\hline
2--1     & 26 & - & 0.37 & 0.74 & - & 0.46 & 0.87 \\
3--2     & 17 & - & 0.54 & 0.61 & - & 0.49 & 0.58 \\
5--4     & 10 & - & 0.77 & -    & - & 0.35 & -    \\
6--5     & 9  & - & 0.39 &  -   & - & 0.17 & -    \\
\hline
\hline
\end{tabular}
\begin{center}
$^{\rm (a)}$ Obtained from gaussian fits (see Sect. 4). $^{\rm (b)}$ At the original HPBW. $^{\rm (c)}$ Estimated in a 20$''$ HPBW.
\end{center}
\end{table}

\section{Physical conditions of the gas}

The fluxes used for this analysis are determined as follows.

First, we use the $g_2$-dominated CS (7--6) line profile to obtain the $g_2$ contribution to the low$-J$ CS lines. To produce the $g_2$ profiles we scaled the CS (7--6) high-velocity intensity to the intensity of the high-velocity emission of the low$-J$ CS transitions. This method implicitly assumes optically thin emission (as indeed supported by our analysis in Sect. 3.1) and therefore that the line shape is the same for all the transitions.

Second, the $g_3$ spectra were subsequently produced by subtracting the $g_2$ spectra from the original low$-J$ CS profiles (i.e. the $g_3$ component is the residual). The $g_2$ and $g_3$ contributions of the C$^{34}$S low$-J$ lines were produced in the same way, i.e. scaling to the CS (7--6) profile.

As said above (sect. 3.2), the HIFI CS profiles do not have sufficient signal-to-noise ratio to reliable separate them into exponential components. We then take their total integrated flux from gaussian fits to their profiles. As discussed in Sect. 4.3, it appears that these transitions arise from a region of much higher excitation than $g_2$ and $g_3$, which we tentatively identify with the jet impact shock region, i.e. component $g_1$  in the terminology of Lefloch et al. (2012). The integrated fluxes of the different components $g_1$, $g_2$ and $g_3$ are reported in Table \ref{tab:flux-g2g3}. 

We analysed the excitation conditions of the CS line emission with a radiative transfer code in the LVG approach by using the code described in \citet{Cecclvg}. We employed the He collisional rates from Lique et al. (2006), and considered 31 levels in the calculations. The model includes the effects of beam filling factor (assuming a source size), and it computes the reduced chi-square ($\chi^2_{\rm r}$) for each column density, minimizing with respect to kinetic temperature ($T_{\rm kin}$), and H$_2$ density ($n_{\rm H_2}$).
We adopted a size of $18\arcsec$ for $g_2$, as derived from our PdBI map of the CS 3--2 emission, and a typical size of $25\arcsec$ for $g_3$. These values are consistent with the sizes estimated from our previous analysis of the CO cavities (Lefloch et al. 2012).
The average FWHM linewidths are 7.0, 7.5, and 5.4 km s$^{-1}$, for $g_1$, $g_2$, and $g_3$, respectively, and are also used as input in the calculations. The error in the flux measurements considered in the calculations are, taking into account both statistical and calibration uncertainties, about 20\% for the low$-J$ CS lines, and up to 30\% for the HIFI lines (dominated by noise). 

In order to account for beam coupling effects due to the different beam sizes of each transition, we convolved the IRAM-30m CS (3--2) map to different beam sizes and apply a beam coupling correction factor to the measured fluxes of each component. Since the HPBW of the map is 17$''$, the beam coupling correction obtained by using the IRAM-30m map can only be applied to the lines with beam sizes larger than this value. The change in the line intensities by convolving the IRAM-30m CS (3--2) map to a 20$''$ beam is shown in Fig. \ref{coupling}: the ratio between the original and the convolved spectra seems to slightly depend on the velocity, suggesting slightly different emitting sizes (see below). We used the ratios to correct the intensities and to consequently obtain a set of fluxes of the $g_2$ and $g_3$ components corrected by beam-coupling effects. The correction factors were taken at $+$1 km s$^{-1}$ for $g_3$ (i.e. at a typical velocity where $g_3$ emission is brighter than the $g_2$ one) and at $-$2 km s$^{-1}$ for $g_2$. On the other hand, for the mid$-J$ CS lines, i.e., $J$ = 5--4, 6--5, and 7--6 lines, whose beam sizes are smaller than 17$''$, we used the interferometric CS (3--2) map (synthesized beam $\sim$ 3$''$), which trace pure $g_2$ emission (see Fig. 6), to correct their $g_2$ fluxes from beam coupling effects. To be consistent with the method applied by Lefloch et al. (2012), we used the line intensities as measured in a beam of 20$''$. The fluxes of each component with and without the coupling factor corrections are shown in Table \ref{tab:flux-g2g3}.

The solutions for the LVG analysis are presented in Fig. \ref{lvg-g2g3} as $\chi ^2$ plots in the $T_{\rm kin}$ versus $n$ plane. These solutions are discussed here for each component, and the physical conditions constrained by the CS lines are summarized in Table \ref{tab:lvg}.

   \begin{figure}
   \centering
      \includegraphics[angle=0,width=8.5cm]{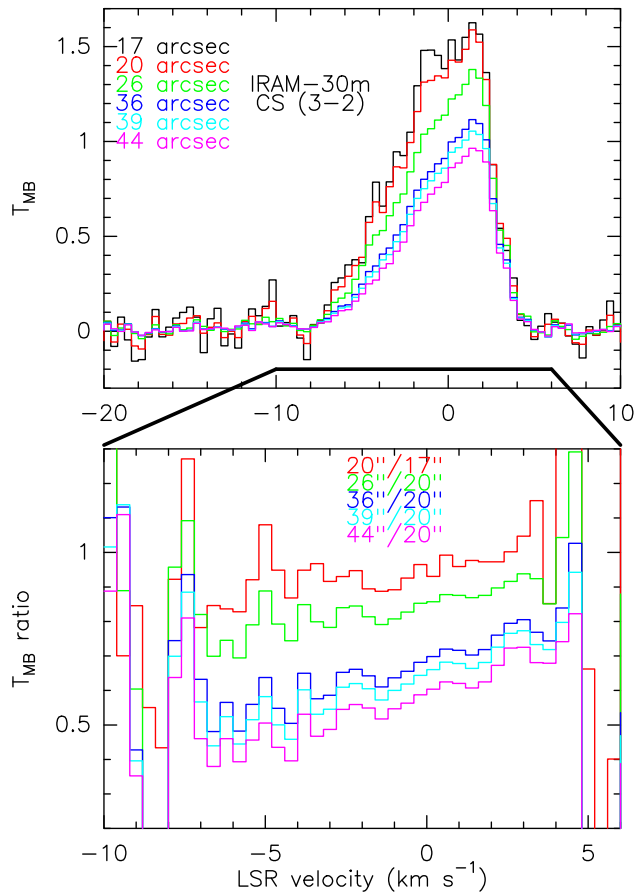}
      \caption{Upper: The resulting CS (3--2) spectra after the convolution of the original IRAM-30m map to different beam sizes. The thick black line shows the velocity range of the lower panel. Lower: The T$_{\rm MB}$ line ratios used to correct for beam coupling in a beam of 20$''$. The beam sizes run from smallest at the top to largest at the bottom.
              }
         \label{coupling}
   \end{figure}

\begin{table}
\centering
\caption{Physical parameters constrained by the CS lines.}
\label{tab:lvg}
\centering
\begin{tabular}{lccc}
\hline
\hline
Component & Size$^{\rm (a)}$ & N(CS)       & $n_{\rm H_2}$  \\
          &  ($''$)  & (cm$^{-2}$) & (cm$^{-3}$)\\
\hline
$g_1$     & 10   & 1 $\times$10$^{13}$    & $>$ 10$^6$\\
$g_2$     & 18   & 8 $\times$10$^{13}$ &  10$^5$--10$^6$\\
$g_3$     & 25   & 8 $\times$10$^{13}$ & 0.5--2 $\times$ 10$^{5}$\\
\hline
\hline
\end{tabular}
\begin{center}
{$^{\rm (a)}$ Based on the CS (3--2) interferometric map and CO results (see sect. 3.3 and 4.3).}
\end{center}
\end{table}

\subsection{The $g_3$ (L1157-B2) cavity}

The $g_3$ component was analysed using the $J$ = 2--1 and 3--2 transitions of the CS and C$^{34}$S isotopologues. We find valid solutions (i.e. $\chi^2_{\rm r}$ = 1) for source sizes larger than 20$\arcsec$. Adopting a source size of 25$\arcsec$ (see Sect.3.3), the total CS column density is then  $8\times 10^{13}$ cm$^{-2}$, and an H$_2$ volume density in the range of 0.5 -- 2 $\times$10$^5$ cm$^{-3}$ (see Fig. \ref{lvg-g2g3}). Taking into account the temperature range constrained by CO, 20--30 K (best-fit: 23 K; Lefloch et al. 2012), the H$_2$ volume density of $g_3$ can be further constrained to $\sim$ 1 $\times$10$^5$ cm$^{-3}$. Using the source-averaged column density $N$(CO) = 1 $\times$ 10$^{17}$ cm$^{-2}$ (found for $g_3$ by Lefloch et al. 2012), and assuming [CO]/[H$_2$]=10$^{-4}$, we can derive the CS abundance: $X$(CS) $\simeq$ 8 $\times$ 10$^{-8}$. This value (i) is in agreement with the abundance estimate by Bachiller (2001) and Tafalla et al. (2010), who derived $\sim$ 10$^{-7}$, using the LTE approximation, for the low-velocity regime (--3.5 $\leq$ $V$ $\leq$ +0.5 km s$^{-1}$) of the low--$J$ CS emission towards L1157-B1, and (ii) is about one order of magnitude larger than what quoted by Bachiller et al. (2001) towards the position of the driving protostar L1157-mm (3 $\times$ 10$^{-9}$), thus confirming that the $X$(CS) can increase in the shocked material located along the cavities.

In addition, we can derive a rough measure of the excitation temperatures from the ($T_{\rm ex}$-$T_{\rm bg}$) $\times$ ff product derived from the $^{12}$CS and $^{13}$CS $J$ = 2--1 spectra and plotted in Fig. 4. With the assumed size of 25$\arcsec$, and consequently correcting the $J$ = 2--1 emission for the filling factor ff, we have $T_{\rm ex}$ $\simeq$ 7--8 K in the --2,0 km s$^{-1}$ range; in other words we find a sub-thermal regime ($T_{\rm kin}$ $\sim$ 23 K), consistently with a $J$ = 2--1 critical density $\simeq$ 10$^6$ cm$^{-3}$, which is slightly larger than what found in the LVG analysis (few 10$^5$ cm$^{-3}$).

\subsection{The $g_2$ (L1157-B1) cavity}

We have first attempted to model the emission of {\em all} the lines from $J$ = 2 to 12, after subtracting the $g_3$ contribution. We could not find any set of physical conditions (i.e. no valid $\chi ^2$) that accounts simultaneously for the low--$J$ CS lines detected with IRAM and those detected with \emph{Herschel} ($J$ = 10--9, 11--10, and 12--11). As we show hereafter, we found however that the separate analysis of the IRAM and \emph{Herschel} lines leads to solutions fully consistent with the previous observations of the region.

We have then modelled the low-- and mid--$J$ lines detected at millimeter and submillimeter wavelengths, adopting a typical size of 18$\arcsec$ as derived from the interferometric map of the $J$ = 3--2 emission. We find column densities of $\sim$8 $\times$ 10$^{13}$ cm$^{-2}$ and H$_2$ volume densities in the range of $10^5$--10$^6$ cm$^{-3}$ (see Fig. \ref{lvg-g2g3}). These results are fully consistent with the physical conditions in $g_2$, as derived from CO (Lefloch et al. 2012). Figure \ref {lvg-g2-sed} reports the best-fit for the spectral line energy distribution for the C$^{32}$S and C$^{34}$S emission of $g_2$ (red line). Taking into account the temperature range constrained by CO, 60--80 K (best-fit: 63 K; Lefloch et al. 2012), the density range from CS can be further constrained to $\sim$ 1--5$\times$10$^5$ cm$^{-3}$. Using again the source-averaged CO column density ($\sim$ 10$^{17}$ cm$^{-2}$), we find a $X$(CS) of 8 $\times$ 10$^{-8}$ (i.e. similar to that found towards $g_3$).

We notice, in addition, that the CS column densities and H$_2$ volume densities for this component is within the range of values reported, based on the analysis of the interferometric CS (2--1) and (3--2) maps, by Benedettini et al. (2013) for some of the high-velocity clumps. In particular, the high-velocity clumps B1a and B1b (both within our beam) are reported with  N(CS)$=$ 2--8$\times$10$^{13}$ cm$^{-2}$ and $n_{\rm H_2}=$ 0.2--5$\times$10$^5$ cm$^{-3}$ (Table 3 in Benedettini et al. 2013). The similar physical conditions and the fact that the interferometric data alone trace mostly the $g_2$ component (see Fig. \ref{log-pdb-30m}) suggest that at least the high-velocity part of the $g_2$ emission is related with the B1a and B1b clumps. 

\subsection{The High-Excitation CS emission}

We analysed separately the flux of the transitions detected with HIFI. We have adopted a typical size of $\sim 10\arcsec$ for the emitting region, bearing in mind that the actual size does not influence the density and temperature derived from the LVG modelling provided that the lines are optically thin. Our LVG calculations yield column densities of $\sim$ 10$^{13}$ cm$^{-2}$ and H$_2$ volume densities $n_{\rm H_2} > 1\times 10^6\cmmt$ (see Fig. \ref{lvg-g2g3}), and indicates that the three transitions are optically thin.

The analysis of the CO and H$_2$O line emission in L1157-B1 (Lefloch et al. 2012; Busquet et al. 2014) has shown evidence for a region (with a typical size of $\sim 10\arcsec$) of dense and hot gas ($\rm n_{\rm H_2} > 10^6$ cm$^{-3}$, $\rm T_k \simeq 250$~K) associated with the impact of the protostellar jet against the B1 cavity. Based on the similarity of the physical conditions, we propose that the high-J CS lines are tracing the jet shock impact region.

Taking into account the constrain on the $g_1$ kinetic temperature given by the CO and H$_2$O observations (Lefloch et al. 2012; Busquet et al. 2014) of 200--300 K (best-fit: 210 K), the volume density of the $g_1$ component can be further constrained to 1--5$\times 10^6\cmmt$. Figure \ref {lvg-g2-sed} reports the best-fit for the spectral line energy distribution for the $g_1$ emission (blue line). According to this model, the integrated low$-J$ CS emission from $g_1$ would be at least an order of magnitude smaller than the error bars reported for the total CS emission, and two orders of magnitude weaker than the combined $g_2$ and $g_3$ contribution. The integrated low$-J$ CS emission from $g_1$ would then be hidden by the stronger $g_2$ and $g_3$ contributions in our data. Also, using the predicted CS (2--1) integrated intensity of $g_1$, $\sim$ 121 mK (Fig. 10), and asumming the intensity-velocity relatioship $\rm \propto exp(-|v/12.5|)$ found in CO for the $g_1$ component (Lefloch et al. 2012), we can predict the $g_1$ intensity at the velocities where the $g_2$ and $g_3$ emission is not significant in the CS (2--1) spectrum, i.e. $\rm v < - 20$ km s$^{-1}$. With the relationship obtained, $\rm I(v) = 0.012895 * exp(-|v/12.5|)$, at $\rm v = - 20$ km s$^{-1}$ the expected intensity is $\sim$ 2.6 mK (hence $<$3$\sigma$ detection limit; see Table 2), making the $g_1$ component undetectable for $\rm v \leq - 20$ km s$^{-1}$ even in the strongest low$-J$ CS lines. 

Comparing with $N$(CO) = 1 $\times$ 10$^{16}$ cm$^{-2}$, we derive the CS abundance in $g_1$ $X$(CS) $\simeq$ 1 $\times$ 10$^{-7}$, which is not very different from the values found for $g_2$ and $g_3$. The overall $X$(CS) measurements are consistent with the values found by Tafalla et al. (2010) who analysed the molecular content of two prototypical jet-driven outflows such as L1448 and IRAS 04166+2706, reporting a velocity-dependent CS abundance which increases at most one order of magnitude. In addition, we compared the inferred CS abundances with the prediction made by Podio et al. (2014) by computing the chemical evolution of the gas (from steady-state abundances to enhanced values produced by compression/heating of a shock wave) at the density and temperature of the cavities in L1157-B1. To model the emission of several molecular ions, these authors assumed that the OCS is released from the dust icy mantles due to dust grain sputtering. Their observations are matched by a model in which at the shock age $\sim$ 2000--4000 yr the X(OCS) is enhanced up to values $\geq$ 2$\times$10$^{-6}$. For their best-fit model (X(OCS) = 6$\times$10$^{-6}$), the abundance of CS is $\sim$ 10$^{-7}$. This value is in good agreement with the overall CS abundances inferred here.


  \begin{figure}
   \centering
    \includegraphics[bb=81 41 556 547,angle=-90,width=8.0cm]{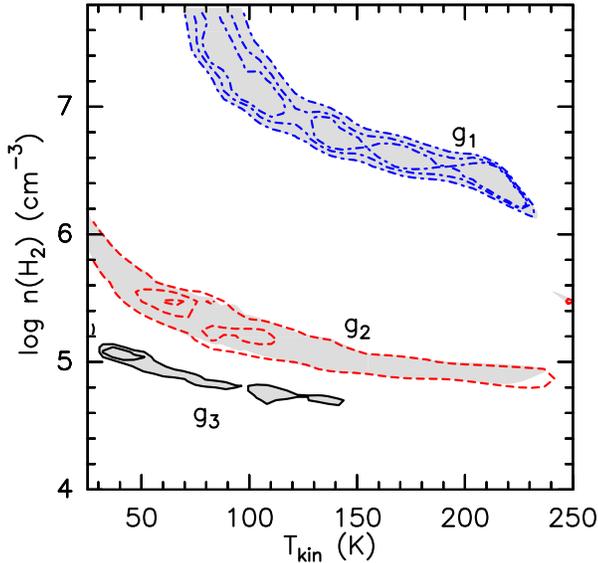}
      \caption{The $\chi^2_{\rm r}$ distribution for the $g_1$, $g_2$, and $g_3$ components (dot-dashed blue, dashed red, and black line, respectively). The valid solution region is delimited by the contour $\chi^2_{\rm r}$ =1. For $g_1$ and $g_2$, $\chi^2_{\rm r}$ contours are 1, 0.5, and 0.2; while for $g_3$ are 1 and 0.83 (min $\chi^2_{\rm r}$). }
         \label{lvg-g2g3}
   \end{figure}

   \begin{figure}
   \centering
      \includegraphics[angle=0,width=9.0cm]{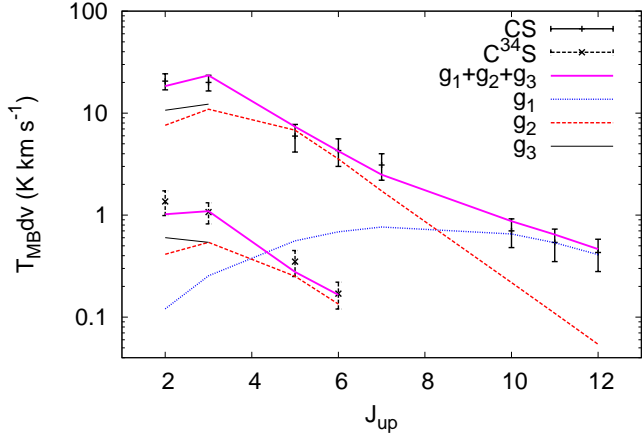}
      \caption{
The CS and C$^{34}$S spectral line energy distribution (data points with error bars) and a three components model with the $g_1$ (dot-dashed blue line), $g_2$ (dashed red line), and $g_3$ (black line) components. The sum of the components is indicated by the thick magenta line. For $g_3$, assuming a size of 25$\arcsec$ (Lefloch et al. 2012), the best fit gives N(CS) = 8 $\times$ 10$^{13}$ cm$^{-2}$, $n_{\rm H_2}$ = 2$\times$10$^{5}$ cm$^{-3}$, and $T_{\rm kin}$ = 30 K. For $g_2$, assuming a size of 18$\arcsec$ (see Fig. 1), the best fit gives N(CS) = 8 $\times$ 10$^{13}$ cm$^{-2}$, $n_{\rm H_2}$  = 2$\times$10$^{5}$ cm$^{-3}$, and $T_{\rm kin}$ = 90 K. For $g_1$, assuming a size of 10$\arcsec$ (Lefloch et al. 2012, Busquet et al. 2014), the best fit reports N(CS) = 1 $\times$ 10$^{13}$ cm$^{-2}$, $n_{\rm H_2}$ = 2 $\times$ 10$^{7}$ cm$^{-3}$, and $T_{\rm kin}$ = 90 K.
              }
         \label{lvg-g2-sed}
   \end{figure}

\section{Summary and conclusions}

In this paper we have presented a CS multiline study based on data obtained with \emph{Herschel}-HIFI and IRAM-30m at L1157-B1, within the framework of the CHESS and ASAI surveys. The main results are summarized as follows:

\begin{enumerate}

\item We have detected $^{12}$C$^{32}$S, $^{12}$C$^{34}$S, $^{13}$C$^{32}$S, for a total of 18 transitions, with $E_{\rm u}$ up to $\sim$ 180 K. The unprecedented sensitivity of the survey allows us to carefully analyse the line profiles, revealing high-velocity emission, up to 20 km s$^{-1}$ with respect to the systemic velocity. With the use of the isotopologues we confirmed that the emission is optically thin at the outflow velocities ($\tau$$\sim$0.05 at $-$7.5 km s$^{-1}$, while $\tau$$\sim$1 at the cloud velocity).

\item The profiles can be well fitted by a combination of two exponential laws that are remarkably similar to what previously found using CO. These components have been related to the cavity walls produced by the $\sim$ 2000 yr B1 shock (called $g_2$) and the older ($\sim$ 4000 yr) B2 shock ($g_3$), respectively. Previous CO observations allowed us to derive the kinetic temperatures, i.e. 23 K and 64 K for $g_3$ and $g_2$, respectively. Using the LVG approximation, we can now put severe constrains on volume density: both the B1 and B2 large cavities are associated with $n_{\rm H_2}$ $\simeq$ 1--5 $\times$ 10$^{5}$ cm$^{-3}$. In addition, the high-excitation (E$_u$ $\geq$ 130 K) CS lines provide us with the signature of warm ($\sim$ 200--300 K) and dense ($n_{\rm H_2}$  = 1--5 $\times$ 10$^{6}$ cm$^{-3}$) gas, associated with a molecular reformation zone of a dissociative J-type shock (previously detected using [OI], [FeII], CO and H$_2$O)  and expected to arise where the so far unrevealed precessing jet impacts the molecular cavity.

\item Our analysis confirms that the CS abundance in shocks increase up to 0.8--1 $\times$ 10$^{-7}$, i.e. more than one order of magnitude with respect to what is found in the hosting cloud, in agreement with the prediction of the model obtained via the chemical code Astrochem. Such enhancement is possibly due to the release of OCS from dust grain mantles, as suggested by Wakelam et al. (2004), Codella et al. (2005), and, more recently by Podio et al. (2014).

\end{enumerate}

\section{Acknowledgments}
We thank the anonymous referee for the detailed comments which helped to improve the clarity of this paper. We are grateful to S. Cabrit for useful discussion and suggestions. The Italian authors gratefully acknowledge the support from the Italian Space Agency (ASI) through the contract I/005/011/0, which also provided the fellowships of A.I G\'omez-Ruiz and G. Busquet. G.B is supported by the Spanish MICINN grant AYA2011-30228-C03-01 (co-funded with FEDER funds). HIFI has been designed and built by a consortium of institutes and university departments from across Europe, Canada and the United States under the leadership of SRON Netherlands Institute for Space Research, Groningen, The Netherlands and with major contributions from Germany, France and the US. Consortium members are: Canada: CSA, U. Waterloo; France: CESR, LAB, LERMA, IRAM; Germany: KOSMA, MPIfR, MPS; Ireland, NUI Maynooth; Italy: ASI, IFSI-INAF, Osservatorio Astrofisico di Arcetri-INAF; The Netherlands: SRON, TUD; Poland: CAMK, CBK; Spain: Observatorio Astron\'omico Nacional (IGN), Centro de Astrobiolog\'ia (CSIC-INTA). Sweden: Chalmers University of Technology - MC2, RSS \& GARD; Onsala Space Observatory; Swedish National Space Board, Stockholm University - Stockholm Observatory; Switzerland: ETH Zurich, FHNW; USA: Caltech, JPL, NHSC.


\end{document}